\documentstyle[twocolumn,aps]{revtex}
\begin{document}

\draft
\preprint{IUCM95-013}
\title{Skyrme Crystal In A Two-Dimensional Electron Gas}

\author{L. Brey$^1$ H.A.Fertig$^2$, R.C\^ot\'e$^3$ and A.H.MacDonald$^4$.}
\address{$^1$Instituto de Ciencia de Materiales (CSIC).
Universidad Aut\'onoma C-12,
28049, Madrid, Spain.  }
\address{$^2$ Department of Physics and Astronomy, University of Kentucky,
Lexington, Kentucky 40506-0055.}
\address{$^3$ D\'epartement de Physique et Centre de Recherches en Physique du
Solide, Universit\'e de Sherbrooke, Sherbrooke, Qu\'ebec, Canada J1K 2R1.}
\address{$^4$ Department of Physics, Indiana University,
Bloomington, Indiana 47405.}

\date{\today}

\maketitle

\begin{abstract}

The ground state of a two-dimensional electron gas at
Landau level filling factors near $\nu =1$ is a Skyrme crystal
with long range order in the positions and orientations of
the topologically and electrically charged elementary
excitations of the $\nu=1$ ferromagnetic ground state.
The lowest energy Skyrme crystal is a square lattice
with opposing postures for topological excitations on
opposite sublattices.  The filling factor dependence of the
electron spin-polarization, calculated for the square lattice
Skyrme crystal, is in excellent agreement with recent experiments.

\end{abstract}

PACS number 73.20.Dx


The incompressible\cite{hall} ground state of a two-dimensional
electron gas (2DEG) at Landau level filling factor $\nu =1$ is
ferromagnetic.  (Here $\nu \equiv N/N_{\phi}$ is the ratio of
the number of electrons to the orbital degeneracy of a Landau level;
$N_{\phi} = A B /\Phi_0 = A / (2 \pi \ell^2)$ where $A$ is the
area of the system, $\Phi_0$ is the magnetic flux quantum, and B is the
magnetic field strength.)
Recently it has been shown\cite{sondhi,rezayi,lee,moon} that, in
the limit of weak Zeeman coupling,
the lowest energy charged excitations of this state are spin-textures
known as {\it skyrmions}.\cite{rajaraman}.
Skyrmions are the lowest energy topologically charged
spin-texture excitations of the SU(2)
non-linear $\sigma$ (NL$\sigma$) model which describes the long wavelength
properties of isotropic ferromagnets.
The equivalence of physical charge and topological charge
in the present system is a consequence\cite{sondhi,moon}
of the quantum Hall effect\cite{hall} and is responsible for the dominating
role of skyrmions in determining many physical properties.

Pioneering studies of skyrmions in the quantum Hall regime relied on
NL$\sigma$ models generalized\cite{sondhi,moon} to account for Zeeman coupling
to the spin and for the Hartree (electrostatic) interactions present when
the charge density is non-uniform.
States in the NL$\sigma$ model are specified by a space-dependent unit
vector which describes the local orientation of the spin magnetic moment.
A skyrmion is characterized by the sign of its topological charge,
by its size, and by the global
orientation of the spin; with no Zeeman or Hartree coupling
the energy of an isolated skyrmion is independent of all three.
When Zeeman coupling is included the spin moment
outside a skyrmion aligns with the field ($m_z =1$),
the spin moment at the center of a skyrmion is oriented in opposition
to the field ($m_z =-1$).  The perpendicular component ($m_{\perp}$) of the
spin-moment, which
must be non-zero as $m_z$ changes from $-1$ to $1$, has a vortex at the
skyrmion center.
The global azimuthal orientation of $m_{\perp}$
is, importantly for the
work described here, still arbitrary.
Because the spin-moment is reversed in the interior of a skyrmion, Zeeman
coupling favors small
skyrmions.  On the other hand the Hartree self-interaction energy of the
skyrmion favors large
skyrmions so that an optimal skyrmion size is established.\cite{sondhi}
For Zeeman coupling strengths typical of physical systems the
estimated skyrmion size is comparable to microscopic lengths, motivating
a microscopic approach.  Recently\cite{fertig}, using a microscopic
Hartree-Fock approximation, we confirmed the main predictions of
the field-theory approach and obtained an estimate of the number of reversed
spins for each (skyrmion) charge added to the system at $\nu =1$.
The quantitative agreement between this estimate and subsequent
Knight shift measurements\cite{exp} appears to provide incontrovertible
evidence in favor of the exotic topological nature of the low-energy charge
carriers in this system.  In this Letter we report on a Hartree-Fock
approximation
study of the ground state of the 2DEG for $\nu$ near $1$, when interactions
between the charged spin-textures, which we will refer to as skyrmions
even when distorted by Zeeman and Hartree couplings, become important.

The most important result of our calculation is shown in
Fig.~\ref{fig:1} where we compare our theoretical results for the
spin-polarization $P$ of the two-dimensional electron gas as a function
of filling factor with experiment.\cite{exp}  ($P$ is the
spatial average of $m_z$.) Without interactions
the spins are completely aligned by the Zeeman field for $\nu < 1$.
The polarization is gradually reduced for $\nu > 1$, reaching zero for
$\nu =2$, since the
Pauli exclusion principle forces the occupation of reversed spins.
The large slope of $P$ on either side
of $\nu=1$ is in sharp contrast with the non-interacting electron
result and can be related to the number of reversed spins in an
isolated skyrmion.\cite{exp,sondhi,fertig}
Our calculations at finite $|\nu -1|$ were
performed assuming that the ground state of the system is a
crystal of charged skyrmion spin-textures.  These states are analogous to the
Skyrme crystal states which arise\cite{nuclear} in studies of
dense nuclear matter using Skyrme's topological excitation model.
The Hartree-Fock equations have many different solutions; all
have long-range order in both the
spatial arrangement of skyrmion centers {\it and} in the
azimuthal orientation of $m_{\perp}$ near each skyrmion center.
Results are shown for two cases: i) a triangular lattice of skyrmions with
identical orientations of $m_{\perp}$ near each site (TLF state)
and, ii) a square lattice with oppositely
directed orientations of $m_{\perp}$ on the two square sublattices
(SLA state).  (The initials A and F in the labels of these
states suggest the obvious analogies with ferromagnetic and
antiferromagnetic states of spin systems on two-dimensional square and
triangular lattices.)
We see in Fig.~\ref{fig:1} that the slope of the
spin-polarization curve decreases in magnitude as we move away from $\nu=1$ in
both cases, indicating
that the skyrmions shrink as they become more dense.  The shrinking
is evidently much more rapid for the TLF state than for
the SLA state.  The results for $P$ for the latter case are in
quite reasonable agreement with experiment and indeed, as we discuss below,
we find that the square Skyrme lattice has lower energy than the
triangular Skyrme lattice except at very small $|\nu -1|$.  Comparison
between theory and experimental results for $P$ suggests that the ground state
is
a square Skyrme crystal until $|\nu -1| $ $\approx 0.2$.

Our calculations are greatly simplified by
working entirely within the lowest orbital Landau level and the
Greens function equation of motion approach we have developed
previously is readily adapted to the present situation.\cite{cote}
For small enough $|\nu-1|$, the skyrmions
can be considered to be point particles and it is known from Madelung
energy calculations for 2D lattices\cite{bonsall}that they
will form a triangular lattice.
Nevertheless, we find that by the time $\nu =1.1$ the
SLA state is lower than the TLF in energy by $\approx 30$ times
the amount by which the Madelung energy would favor the opposite
energetic ordering of the states.
The self-consistent spin-textures for
the TLF and SLA states at $\nu = 1.1$ are illustrated in Fig.~\ref{fig:2}.
The skyrmions are evidently more compact for the higher energy
TLF state and this is the origin of the differences in spin-polarization
between these states.

To understand the differences between SLA and TLF states in
greater depth it is useful to compare the contributions to the
energy per electron from exchange, Hartree and Zeeman energies
as shown in Fig.~\ref{fig:3} for $g \mu_B B = 0.015 e^2/\ell$.
We express our results in terms
of the energy ($\epsilon_{qp}$) increase per quasiparticle, $N_{qp} =
|N-N_{\phi}|$,
when the quasiparticles
are created at fixed $N$ by varying $N_{\phi}$:
\begin{equation}
\epsilon_{qp} \equiv \frac{\nu [\epsilon  +  (e^2/\ell) \sqrt{\frac{\pi}{8}} +
g \mu_B B/2 ]}{|1-\nu|}
\label{eq:eqp}
\end{equation}
where $\epsilon = E/N$ is the energy per electron, $g$ is the g-factor for the
2DEG system,
and $\mu_B$ is the electron Bohr magneton.  (The Hartree, exchange, and Zeeman
contributions to
$\epsilon_{qp}$ are given by terms of corresponding origin on the
right-hand-side of
Eq.~\ref{eq:eqp};$ -(e^2/\ell) \sqrt{\pi/8}$ is the exchange energy
per-particle at $\nu = 1$.)
This convention for the  energy per charge is implicit in the field theoretical
expressions and
we choose it here because it will allow our Hartree-Fock results and results
obtained in the
field-theory description to be compared directly.
In the limit of vanishing skyrmion density
$\epsilon_{qp}$ is the so-called neutral quasiparticle energy.\cite{qpengconv}
It follows from particle-hole symmetry that with this defintion
antiskyrmion quasiparticle energies are smaller than skyrmion
quasiparticle energies by $g \mu_B B$.

We first discuss the constituents of the energy
for the Wigner crystal (WC) state which occurs near $\nu =1$ in the
limit of very strong Zeeman coupling.  In this state a
triangular (Wigner) crystal\cite{macdreview}
of single-reversed-spin quasiparticles is formed on the background of a full
Landau
level of aligned spins.  The reversed spins are localized
to the greatest extent possible with the lowest Landau level.
For the WC state, the exchange and Hartree contributions to the energy
per quasiparticle, $\epsilon_{qp}^X$ and $\epsilon_{qp}^H$,
can be calculated analytically\cite{unpub} in the limit of small
$\nu_{qp}$; $\epsilon_{qp}^X \to (e^2/\ell) [(\pi/8)^{1/2} - (\pi/16)^{1/2}]
\approx 0.1835 (e^2/ \ell)$ and $\epsilon_{qp}^{H} \to
(e^2/\ell) (\pi/16)^{1/2} \approx 0.4431 (e^2/\ell)$,
the Hartree self-interaction energy of a lowest Landau level
quantized cyclotron orbit.  The exchange energy, which is minimized when all
spins are parallel,
 is nearly constant with increasing $\nu_{qp}$
until the quasiparticles begin to overlap significantly while the
Hartree energy is reduced at larger $\nu_{qp}$ due to spatial
correlations of the quasiparticles.  When the size of the quasiparticles
($ \sim \ell$ for the WC state) is much smaller than the distance
between quasiparticles the Hartree energy is reduced by the Madelung
energy per particle\cite{bonsall}; when the
quasiparticles form a triangular lattice $\epsilon_{mad}
\approx -0.7821 \nu_{qp}^{1/2} (e^2 / \ell)$.
In Fig.~\ref{fig:3} we have plotted the difference between the Hartree energy
and the
Madelung energy {\it vs.} $\nu_{qp}$.  For the WC state the remaining variation
is due to the fact that the
electrons are localized in cyclotron orbits and not at points.  The Zeeman
energy
contribution to $\epsilon_{qp}$ is $g \mu_B B \langle S_{qp} \rangle $ where
$\langle S_{qp} \rangle$ is the number of reversed spins per quasiparticle.
($ \langle S_{qp} \rangle $ is smaller for antiskyrmions than for
skyrmions by $1$.)
For the WC state $\langle S_{qp} \rangle \equiv 1$ for skyrmions and
$0$ for antiskyrmions.

In the absence of Zeeman coupling a skyrmion
quasiparticle has lower energy than a reversed spin quasiparticle
because its charge is spread out over an arbitrarily large distance,
reducing $\epsilon_{qp}^H$ to zero while at the same time keeping
spins locally nearly parallel so that $\epsilon_{qp}^X$ increases
only\cite{sondhi,moon}
to $(e^2/\ell) (\pi/32)^{1/2} \approx 0.3133 (e^2/\ell)$.  When
a finite Zeeman coupling is present, the skyrmion quasiparticle
must shrink to reduce its Zeeman energy\cite{sondhi,fertig}
and the Hartree energy increases.  For an isolated skyrmion ($\nu_{qp} \to 0$)
and $g \mu_B B = 0.015 e^2/\ell$ we find using the approach\cite{remark} of
Ref.\onlinecite{fertig} that
$\epsilon_{qp}^H = 0.2499 (e^2/\ell)$, $\epsilon_{qp}^X= 0.2957 (e^2/\ell)$ and
that $\langle S_{qp} \rangle = 3. 83 $.
In Fig.~\ref{fig:3} we see that for the TLF state
the Hartree energy quickly rises from the isolated skyrmion
value toward the WC value and $\langle S_{qp} \rangle$ quickly decreases from
the isolated skyrmion value to  $1$ as $\nu_{qp}$ increases.
Evidently there is a strong short-range repulsive interaction between skyrmions
in the TLF Skyrme state which quickly shrinks them into single-reversed-spin
quasiparticles.
The origin of this repulsion is an increase in exchange energy in the region
between neighboring skyrmions with identical azimuthal orientations of
$m_{\perp}$.
As we see in Fig.~\ref{fig:2} the azimuthal orientation of $m_{\perp}$
must then change by $\pi$ relatively abruptly along the line
connecting neighboring skyrmion centers; in the field-theoretical
description\cite{sondhi,moon} the associated increase in exchange energy is
described by a term $\propto |\nabla m_{\perp}|^2$ and the resulting
repulsive interaction will be roughly proportional to
$m_{\perp}^2$ evaluated at the mid-point between skyrmion centers.
Clearly this repulsive interaction is minimized if neighboring skyrmions have
opposite
azimuthal orientations for $m_{\perp}$ as illustrated for the SLA state in
Fig.~\ref{fig:2}.
(It seems clear that the lowest energy triangular Skyrme lattice will be a
three-sublattice
state with $2 \pi /3$ relative orientation but this state will, we believe, be
higher
in energy than the SLA state unless $\nu$ is very close to 1.)
The exchange energy thus favors bipartite
lattices for a Skyrme crystal.   Our numerical results for the
SLA state show that the skyrmions shrink relatively slowly as $\nu_{qp}$
increases;
for this arrangement the Skyrme crystal is able to
tolerate much tighter packing of the skyrmions and the rapid decrease in
spin-polarization
persists out to quite large values of $\nu_{qp}$ as shown in Fig.~\ref{fig:1}
and
observed experimentally.\cite{exp}

The size of a quasiparticle in a Skyrme crystal is
limited by the lattice constant of the crystal until,
with decreasing $\nu_{qp}$ the lattice constant becomes
substantially larger than the size of an isolated
skyrmion quasiparticle.  For weaker Zeeman couplings,
isolated skyrmions are larger\cite{sondhi,fertig} and spin-polarization will
decrease more rapidly
for $\nu$ very close to $\nu =1$;  for larger values of $|\nu -1|$ where the
skyrmion size is limited
by inter-skyrmion interactions, the spin-polarization is much less dependent on
the Zeeman
coupling.  For example we find for the TLF state at $g \mu_B B / (e^2/ \ell) =
0.0015$
that $P = 0.7 $ at $\nu =1.03$, substantially smaller than the values reported
in
Fig.~\ref{fig:1}, while for $\nu = 1.15 $, $P= 0.7 $ again, nearly identical to
Fig.~\ref{fig:1}.

 \, From our calculations we can conclude that for typical $g \mu_B B$ values,
the ground state of the 2DEG near $\nu=1$ is composed of skyrmionic
charged excitations of the ferromagnetic $\nu=1$ background.  When these
quasiparticles
are sufficiently dilute they will crystallize because of the absence of
kinetic energy.  Although we have not completed an exhaustive survey of
possible Skyrme lattices,
we believe that SLA Skyrme crystal will be lowest in energy
once the lattice constant is
small enough for the gradient energy to overwhelm the relatively
modest structural preferences of the Hartree energy.  It is
difficult for us to estimate the range of filling factors over
which the Skyrme crystal is favored over fluid states of the 2DEG.
However, it seems unlikely that the energy competitions which determines
skyrmion size
will be very sensitive to the existence of long-range-order.
We anticipate the existence of a
Skyrme liquid state, which will be lower in energy than maximally spin
polarized liquid states and
which will have a spin polarization quite close to that of the SLA state.
The Skyrme liquid should have a typical coordination number
closer to four than to six because of the dependence of the exchange
energy on the relative azimuthal orientations of $m_{\perp}$ for nearby
skyrmions.

The authors thank Sean Barrett, Lotfi Belkhir, Steve Girvin,
Peter Littlewood, Aron Pinczuk, Shivaji Sondhi, Carlos Tejedor, and Kun Yang
for helpful discussions.  This work was
supported in part by NATO CRG No. 930684, by the NSF through Grants Nos.
DMR 92-02255 and DMR 94-16906, by the CICyT of Spain under Contract
No. MAT 94-0982, by NSERC of Canada, and by FCAR from the Government of Quebec.
HAF acknowledges the support of the A.P.Sloan
Foundation and the Research Corporation.  LB thanks AT\&T-Bell Labs at
Murray Hill and AHM thanks UAM for hospitality during the realization of this
work.

\begin{figure}
\caption{Variation of the spin polarization, $P$, as a functions of $\nu$, for
SLA and TLF Skyrme crystal states and for different values of $g^{*} \equiv
g \mu_B B/ (e^2/ \ell)$.  The open circles are experimental data obtained for
$g^{*}$ near 0.016, while the filled circles show data obtained for
$g^{*}$ near 0.021. The spin polarization in the WC state is the same as that
for
independent electrons.}
\label{fig:1}
\end{figure}

\begin{figure}
\caption { Two-dimensional vector representation of the $x-y$ components of the
spin
density of a crystal of skyrmions as obtained in the HFA.  The length of the
arrows is
proportional to the local magnitude of $m_{\perp}$ and their direction
indicates the
local orientation.  The magnitude of $m_{\perp}$ is small for $m_z$ near -1,
near
the center of a skyrmion, and for $m_z$ near +1, outside  of a skyrmion.(a) TLF
state at $\nu=1.1$ and $g^{*} = 0.015$; (b) SLA state
at $\nu = 1.1$ and $g^{*} = 0.015$.}
\label{fig:2}
\end{figure}

\begin{figure}
\caption{Zeeman (a) , Hartree (b)  and exchange (c) contributions to the energy
per quasiparticle for the
Wigner crystal state (*) and for the TLF (triangles) and SLA (squares) Skyrme
crystal states. The
results for $\nu =1$ were obtained analytically for the WC state and from
isolated skyrmion
calculations for the Skyrme crystal states.  The Zeeman energy is expressed in
terms of the
average number of reversed spins per quasiparticle and the Madelung
contribution has been
subtracted from the Hartree contribution, as discussed in the text.}
\label{fig:3}
\end{figure}


\begin{references}
\bibitem{remark} Since the $\hat z$-component of the total spin is a
 good quantum number,
for an isolated skyrmion $\langle S_z \rangle$ is an integer in the exact
many-body ground state.
The Hartree-Fock estimate is a continuous function of system parameters and
applies unless $k_B T$ is smaller than energetic separation between
single-skyrmion states with adjacent values of $\langle S_z \rangle$.
For typical
systems we do not expect the discreteness of the allowed values for $\langle
S_z \rangle$ to
be observable above $\sim 1 {\rm K}$.  In the Skyrme crystal skyrmion-skyrmion
interactions
will cause $\langle S_z \rangle$ to fluctuate on each site even at $T=0$.
\end{references}
\end{document}